# TensorSymmetry: a package to get symmetry-adapted tensors disentangling spin-orbit coupling effect and establishing analytical relationship with magnetic order

Rui-Chun Xiao[1,2*], Yuanjun Jin[3], Zhi-Fan Zhang[4], Zi-Hao Feng[1,2], Ding-Fu Shao[5], Mingliang Tian[6,7]

[1]*Institute of Physical Science and Information Technology, Anhui University, Hefei 230601, China*
[2]*Anhui Provincial Key Laboratory of Magnetic Functional Materials and Devices, School of Materials Science and Engineering, Anhui University, Hefei 230601, China*
[3]*Guangdong Basic Research Center of Excellence for Structure and Fundamental Interactions of Matter, Guangdong Provincial Key Laboratory of Quantum Engineering and Quantum Materials, School of Physics, South China Normal University, Guangzhou 510006, China*
[4]*Interdisciplinary Center for Theoretical Physics and Information Sciences (ICTPIS), Fudan University, Shanghai 200433, China*
[5]*Key Laboratory of Materials Physics, Institute of Solid State Physics, Hefei Institutes of Physical Science, Chinese Academy of Sciences, Hefei 230031, China*
[6]*Anhui Key Laboratory of Low-Energy Quantum Materials and Devices, High Magnetic Field Laboratory, HFIPS, Anhui, Chinese Academy of Sciences, Hefei, Anhui 230031, China*
[7]*School of Physics and Optoelectronics Engineering, Anhui University, Hefei 230601, China*
E-mail addresses: xiaoruichun@ahu.edu.cn (R.-C. Xiao)

## Abstract

The symmetry-constrained response tensors on transport, optical, and electromagnetic effects are of central importance in condensed matter physics because they can guide experimental detections and verify theoretical calculations. These tensors encompass various forms, including polar, axial, *i*-type (time-reversal even), and *c*-type (time-reversal odd) matrixes. The commonly used magnetic groups, however, fail to describe the phenomena without the spin-orbit coupling (SOC) effect and cannot build the analytical relationship between magnetic orders with response tensors in magnetic materials. Developing approaches on these two aspects is quite demanding for theory and experiment. In this paper, we use the magnetic group, spin group, and extrinsic parameter method comprehensively to investigate the symmetry-constrained response tensors, then implement the above method in a platform called "TensorSymmetry". With the package, we can get the response tensors disentangling the effect free of SOC and establish the analytical relationship with magnetic order, which provides useful guidance for theoretical and experimental investigation for magnetic materials.

**Program Summary**
*Program title:* TensorSymmetry
*Program Files doi:* http://dx.doi.org/xxxx
*Download*: https://github.com/Ruichun/TensorSymmetryPackage

*Licensing provisions:* GNU General Public Licence 3.0
*Programming language:* Wolfram Mathematica
*External routines/libraries used:* None
*Nature of problem:* To determine the symmetry-adapted tensors in magnetic materials, which can reflect the SOC effect, and establish the analytical relationship with magnetic order.
*Solution method:* Comprehensively using the magnetic groups, spin groups, and the extrinsic parameter method.

## 1. Introduction

Symmetry is a fundamental factor in shaping the physical properties of condensed matter systems. The spontaneous breakings of spatial-inversion symmetry and time-reversal symmetry are two critical phenomena in condensed matter physics, which can lead to ferroelectricity and magnetism, respectively. These symmetry breakings give rise to a wide array of physical phenomena, such as nonlinear optical effect [1-5] (e.g. second harmonic generation (SHG) effect, bulk photovoltaic effect (BPVE), circular photogalvanic effect (CPGE), spin photogalvanic effect (SPGE, also referred to as the spin photovoltaic effect in some contexts)), nonlinear transport effect [6-14], anomalous Hall effect (AHE [15-19], and its optical analog [20] Faraday effect and Kerr effect), spin Hall effect (SHE) [16-19,21], magnetoelectric effect [22], light-induced magnetism [23], Rashba-Edelstein effect (inverse spin galvanic effect) [24] *etc.*. These phenomena not only deepen our understanding of the physics in condensed matter materials but also open avenues for potential technological applications. The study of symmetry-constrained response tensors is crucial in both theoretical and experimental studies, as they govern the experimental detections and validation of theoretical calculations.

Even though there are various kinds of response effects, the corresponding response tensors generally adhere to similar symmetry properties, as shown in Fig. 1. These tensors can be categorized into polar tensors, axial tensors, *T*-even (time reversal even, or *i*-type) tensors, and *T*-odd (time reversal odd, *c*-type) tensors [25]. Typically, Neumann's principle and magnetic point group [26-28] are utilized to investigate the characteristics of the response tensors.

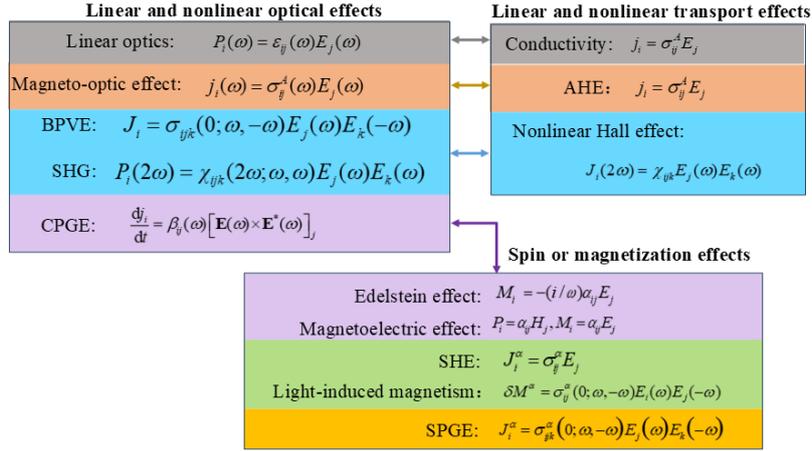

Fig. 1. The expression of linear & nonlinear optical/transport effects, spin, and magnetization effect. The subscript English letters $i, j, k \in (x, y, z)$ mean the direction of current, polarization, or electric field, while the superscript Greek symbols $\alpha \in (S_x, S_y, S_z)$ denote the direction of spin. In SHE and SPGE, $J_i^\alpha$ represents the spin current, where electrons move along the $i$-th direction and their spins are polarized along the $\alpha$-th direction. The double arrow lines among the sections mean the tensors in the response effects share the same symmetry characters.

Among the response effects in the magnetic materials, the spin-orbit coupling (SOC) effect plays a crucial role in SHE, SHG, BPVE, and AHE. In contrast, the *T*-odd SHE [29-31] and *T*-odd SPGE [3,32] are free from the SOC effect. The magnetic group approach has been proven to be inadequate for capturing these SOC-free effects. Refs. [29,30] and [3,32] utilize spin-resolved electron bands in reciprocal space and spin sublattices in real space to elucidate the origins of SHE and SPGE for collinear antiferromagnets, respectively. Even though intuitive, these methods are not convenient and fail for non-collinear magnets [33]. The spin group approach shows a powerful capacity [34,35] to disentangle the SOC effects. As a result, there is a great need for a systematic method to analyze response tensors using the spin group framework.

Furthermore, the response tensors in magnetic materials are strongly dependent on magnetic orders [6,7,36] (such as magnetic moments in colinear ferromagnets, Néel vectors in collinear antiferromagnets, vector chirality in coplanar but noncollinear antiferromagnets). The relationships between the response tensors and magnetic orders hold great significance for detecting magnetic orders [37-39] in experiments, as they provide decoding tables that translate response tensors into magnetic order parameters. However, neither discrete magnetic groups nor spin groups can establish the analytical relationship between these two aspects. Recently, we proposed an "extrinsic parameter" method, which shows a powerful ability to uncover the analytical relationship between the AHE and Néel vectors for altermagnets [40]. This suggests that the approach can be extended to other response effects.

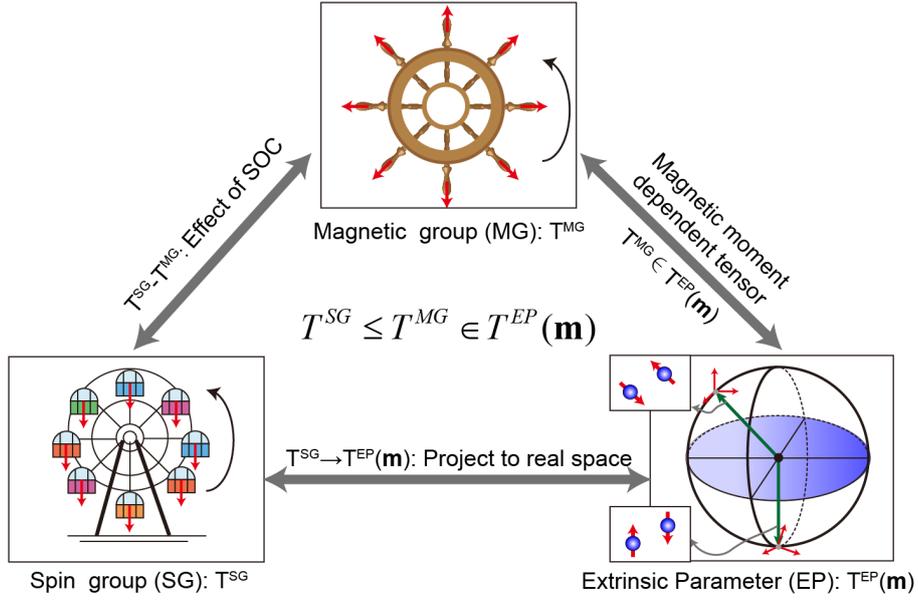

Fig. 2 The integrated platform TensorSymmetry, which combines the magnetic group, spin group, and "extrinsic parameter" methods, gives more comprehensive information about the symmetry-adapted response tensors, to unravel the SOC effect and establish analytical relationships with magnetic orders. In the magnetic group, the spin is coupled with the lattice freedom due to the SOC effect, akin to a steering wheel with fixed handles. Simultaneously, the spin space and real space exhibit partial decoupling in the spin group, a feature reminiscent of a sky wheel, where the orientation of the cabin remains unchanged during rotation. $T^{SG} \leq T^{MG}$ means that the symmetry-adapted tensors constrained by the magnetic group generally have more non-zero tensor elements than those of spin groups, and $T^{MG} \in T^{EP}(\mathbf{m})$ denotes the response tensors under magnetic group belongs to the analytical $T^{EP}(\mathbf{m})$. $\mathbf{m}$ means the magnetic order.

The magnetic group, spin group, and "extrinsic parameter" methods offer valuable insight into response effects. Therefore, a comprehensive platform that integrates all these factors is essential. In this work, we have developed an integrated software platform called "TensorSymmetry" based on the Mathematica language, which combines the three aforementioned methods to analyze symmetry-adapted tensors for magnetic materials, as illustrated in Fig. 2.

The following parts of the paper are arranged as follows: Firstly, we introduce the methods for dealing with the response tensors with the magnetic group, spin group, and extrinsic parameter method in Sec. II-IV. Subsequently, the structure and usage of the software are presented in Sec. V. We will present two representative examples to show the effectiveness of our comprehensive method in Sec. VI.

## 2. Methods

### 2.1 Response tensors constrained by the magnetic groups

Based on whether they change sign under spatial inversion operation, the vectors can be classified into polar vectors (which change signs, such as electric field, current, position, and velocity) and axial vectors (which remain unchanged, such as magnetic field, angular momentum, and spin), respectively. Similarly, response tensors can also be categorized into polar tensors (such as the dielectric tensor, conductivity tensor, and SHG tensor) and axial tensors (such as the magneto-electric tensor, CPGE tensor, and SHE tensor). Additionally, due to the time-reversal symmetry, every response tensor can be further divided into *T*-even (*i*-type) and *T*-odd (*c*-type) parts [25]. Table I summarizes the parities and corresponding mechanisms of SHG, BPVE, CPGE, and nonlinear transport effect.

Table I The parities and corresponding mechanisms of nonlinear optical effect (SHG, BPVE, CPGE) [1,41,42] and nonlinear transport [28,43,44] under the time-reversal symmetry.

|  | *T*-even part (*i*-type) | *T*-odd part (*c*-type) |
|---|---|---|
| SHG | Magnetism & crystal asymmetry | Magnetism |
| BPVE | nonmagnetic shift current | magnetic injection current |
| CPGE | nonmagnetic injection current | magnetic shift current |
| Nonlinear transport effect | Berry curvature dipole | Quantum metric dipole, second order Drude effect |

Generally, Neumann's principle and magnetic point groups are used to investigate the symmetry-adapted response tensors. Based on the parities under the space and time operations, the response tensors can be classified into the following four types [45]:

(1) True tensors (polar & *T*-even): These tensors are transformed under the symmetry operations as:

$$T_{ijk\cdots} = R_{ia}R_{jb}R_{\gamma c}\cdots T_{abc\cdots}. \tag{1}$$

where $i,j,k,a,b,c \in \{x,y,z\}$, $R_{ia}$ is one of the elements of the operation matrix $D(R)$.

(2) Spatial pseudo-tensors (axial & *T*-even). This kind of tensors transform under the symmetry operations as:

$$T_{ijk\cdots} = \det(R)R_{ia}R_{jb}R_{kc}\cdots T_{abc\cdots}. \tag{2}$$

where det(*R*) is the determinant of the operation matrix $D(R)$. For the proper operators (such as C$_1$, C$_2$, C$_3$, C$_4$ and C$_6$), det(*R*)=1, and for improper operator *R* (such as S$_1$, S$_2$, S$_3$, S$_4$ and S$_6$), det(*R*)=-1.

(3) Time pseudo-tensors (polar & *T*-odd): They are transformed under the symmetry operations as:

$$T_{ijk\cdots} = |R|R_{ia}R_{jb}R_{kc}\cdots T_{abc\cdots}. \tag{3}$$

For the unitarity operator (not containing time-reversal operation), $|R|=1$, and for the anti-unitary operator (containing time-reversal operation), $|R|=-1$.

(4) Spatial and time pseudo-tensors (axial & *T*-odd): These tensors transform under the symmetry operations as:

$$T_{ijk\ldots} = \det(R)|R|R_{ia}R_{jb}R_{kc}\cdots T_{abc\ldots}. \tag{4}$$

Besides, the response tensors always have the subscripts-switch symmetry on the electric field. For example, $\chi^{(2)}_{ijk} = \chi^{(2)}_{ikj}$ in the SHG tensor. Therefore, the $\left[\chi^{(2)}_{ijk}\right]_{3\times3\times3}$ tensor can be contracted as a $3\times 6$ matrix $\left[\chi^{(2)}_{il}\right]_{3\times 6}$ using the following indices appointments:

| jk | 11 | 22 | 33 | 23, 32 | 13, 31 | 12, 21 |
|---|---|---|---|---|---|---|
| l  | 1  | 2  | 3  | 4      | 5      | 6      |

Table II The characteristics of 6 common response tensors and corresponding physical effects.

| Response effect | Tensor format | Jahn's notation | Independent tensor elements | Inversion symmetry | PT symmetry | SOC |
|---|---|---|---|---|---|---|
| Conductivity, dielectric function | 2nd order true tensor | [V2] | 6 | √ | √ | √ |
| AHE, Kerr effect, Faraday effect | 2nd order time pseudo-tensor | aVV | 3 | √ | × | × |
| CPGE, Edelstein effect, magnetoelectric effect | T-even: 2nd order spatial pseudo-tensor | eV2 | 9 | × | × | √ |
| | T-odd: 2nd order spatial and time pseudo-tensor | aeV2 | 9 | × | √ | × |
| SHG, BPVE, nonlinear Hall effect | T-even: 3rd order true tensor | V[V2] | 18 | × | × | √ |
| | T-odd: 3rd order time pseudo-tensor | aV[V2] | 18 | × | √ | × |
| SHE, light-induced magnetism | T-even: 3rd order spatial pseudo-tensor | eV3 | 27 | √ | √ | × |
| | T-odd: 3rd order time & spatial pseudo-tensor | aeV3 | 27 | √ | × | √ |
| SPGE | T-even: 4th spatial pseudo-tensor | eVV[V2] | 54 | × | × | × |
| | T-odd: 4th time & spatial pseudo-tensor | aeVV[V2] | 54 | × | √ | √ |

**Note:**

1. In Jahn's notation [27], V means a vector, V2 represents VV, and [] denotes the switch symmetry. The letter "a" represents a c-type (T-odd) tensor, and "e" denotes an axial tensor.
2. The symbols "√" and "×" indicate whether the response effect can persist under the considered symmetry or conditions, respectively.
3. The SOC effect here is for the collinear and coplanar magnetically ordered materials.

The characteristics of transport, optical, and electromagnetic tensors are summarized in Table II, which includes the tensor formats, Jahn's notations [27], and independent tensor elements. Using Eqs. (1)-(4), we can explore the symmetry-constrained tensors under all the magnetic groups. This theoretical framework is straightforward to implement in computational programs, such as the web-based MTENSOR [27] (https://www.cryst.ehu.es/cgi-bin/cryst/programs/mtensor.pl) and the offline "linear-response-symmetry" software [46] (https://bitbucket.org/zeleznyj/linear-response-symmetry/src/master/). We also provided a dictionary on dielectric tensors, AHE tensors, SHG tensors, SHE tensors, and SPGE tensors for all magnetic point groups, which is available on the website: https://ruichun.github.io/TensorSymmetry/. In addition, we recently utilized the isomorphic group method [36] to streamline and enhance the process of deriving the response tensors, making it both more efficient and conceptually meaningful.

The above $T$-odd response effects are attributed to magnetism because the spin is odd udder the time-reversal symmetry. Furthermore, we found that the SOC effect is indispensable for the $T$-odd SHG [36,47] and AHE [19,48] in collinear and coplanar magnetic materials. Conversely, for other response effects, such as $T$-odd SHE and SPGE, the SOC effect is not pivotal. In the following, we will analyze these issues using the spin group method.

**2.2 Response tensors constrained by the spin groups**

The response tensors free of the SOC effect can be investigated with the spin groups [49-52]. Furthermore, the combination of spin group and magnetic group methods can disentangle the influence of the SOC effect.

Due to the partial decoupling between spin space and real space in spin groups as illustrated in Fig. 2, we need to employ two relatively independent sets of symmetry operations, acting on spin and lattice degrees of freedom separately. If $R$ denotes the point symmetry operation in real space, and $U$ represents the symmetry operation in the spin space, then a specific spin point group operation can be written as $\{U \parallel R\}$, and it is defined as $^{U}R$ in international symbol notation [50,52]. For spin group operations, we conventionally treat spin as a polar vector rather than an axial vector to include the time-reversal operation $T$ in spin space. In this context, the time-reversal operation $T$ on spin is equivalent to performing a spatial inversion operation on a polar vector, *i.e.* $T = \bar{1}$. The number of spin point groups (598) is bigger than that of magnetic point groups (122) [50]. Besides, for coplanar and collinear magnets, their spin point groups are the direct product of nontrivial spin point groups and spin-only groups ($m$ and $\infty m$ for coplanar and collinear magnets, respectively). Specifically, 252 and 90 spin point groups are found to describe the coplanar and collinear magnetic structures [52], respectively.

In the following, we will analyze the response tensors free of SOC using the spin group method.

*AHE tensors under the spin groups*. The AHE arises from the transverse motion of electrons under the electric field in real space and is not directly related to the electron's spin. Therefore, the constraint imposed by spin point group operation $\{U \| R\}$ on the AHE tensor (the antisymmetric tensor of conductivity) $\left[\sigma_{ij}^A\right]$ is given by:

$$\{U \| R\}: \sigma_{ij}^A = \det(U) \sum_{mn} R_{im} R_{jn} \sigma_{mn}^A. \tag{5}$$

In the above equation, det($U$) represents the determinant of the spin operation matrix $D(U)$, as AHE changes sign under time-reversal operation. The collinear or coplanar magnets possess pure spin operation $\{m\|E\}$, where $m$ means the mirror symmetry in spin space, and $E$ denotes the identity operation in lattice space. According to Eq. (5), det($m$)=-1 leads to $\sigma_{ij}^A = -\sigma_{ij}^A$. Therefore, spin group point operation $\{m\|E\}$ vanishes the AHE when the SOC effect is unconsidered for the collinear and coplanar magnetic materials, even though they have non-zero net magnetic moments. Conversely, for non-coplanar magnetic materials where the $\{m\|E\}$ symmetry is broken, the AHE and magneto-optical effects can arise even in the absence of the SOC effect [19].

*SHG tensors under the spin groups.* As mentioned, the SHG effect and other nonlinear optical and transport effects involve two types of tensors based on their parity under time-reversal operations: the *T*-even (*i*-type) $\chi_{ijk}^{\text{even}}$ and *T*-odd (*c*-type) $\chi_{ijk}^{\text{odd}}$ tensors (Table I). Under the spin point group operation $\{U \| R\}$, $\chi_{ijk}^{\text{even}}$ (*i*-type) transforms as:

$$\{U \| R\}: \chi_{ijk}^{\text{even}} = \sum_{lmn} R_{il} R_{jm} R_{kn} \chi_{lmn}^{\text{even}}. \tag{6}$$

Differently, $\chi_{ijk}^{\text{odd}}$ transforms as [34]:

$$\{U \| R\}: \chi_{ijk}^{\text{odd}} = \det(U) \sum_{lmn} R_{il} R_{jm} R_{kn} \chi_{lmn}^{\text{odd}}. \tag{7}$$

According to Eq. (7), the pure spin operation $\{m\|E\}$ in magnetically ordered materials with collinear or coplanar structures results in det($m$)=-1, causing $\chi_{ijk}^{\text{odd}}$ vanishing. This explains why the observation of *c*-type (*T*-odd) SHG and BPVE [36,47] in collinear and coplanar magnetic materials requires the SOC effect. Furthermore, the stronger the SOC effect in the collinear or coplanar magnetic materials, the larger the SHG coefficients [36,41,47,53].

*SHE tensors under the spin groups.* Since the SHE involves the motion of both spin and real space, we need to impose two kinds of symmetry constraints on the corresponding indices of SHE coefficients $\sigma_{ij}^{\alpha}$ (the superscript Greek letter denotes the component of spin space, and the subscript English letters represent the components of the real space). The transformation of *T*-even $\sigma_{ij}^{\alpha(e)}$ (*e* is short for even) under the spin point group operation $\{U\| R\}$ is as follows:

$$\{U \| R\}: \sigma_{jk}^{\alpha(e)} = \det(U)\sum_{lmn} U_{\alpha\beta} R_{jm} R_{kn} \sigma_{mn}^{\beta(e)}. \tag{8}$$

Collinear magnetic materials possess a pure spin-only group $\infty m$ [52]. According to Eq. (8), the pure spin operation $\{m_{001}\| E\}$ leads to the vanishing of SHE tensors $\sigma^{S_x(e)}$ and $\sigma^{S_y(e)}$. Additionally, the pure spin mirror operation $\{m_{100}\| E\}$ make all the SHE components of $\sigma^{S_z(e)}$ be zero. Therefore, the *T*-even SHE tensor is zero in the spin group, i.e., *T*-even SHE vanishes in the absence of the SOC effect for collinear magnetic materials. On the contrary, the transformation of the *T*-odd SHE tensor $\sigma_{jk}^{\alpha(o)}$ (*o* is short for odd) under spin point operations $\{U\| R\}$ is:

$$\{U \| R\}: \sigma_{jk}^{\alpha(o)} = \sum_{\beta mn} U_{\alpha\beta} R_{jm} R_{kn} \sigma_{mn}^{\beta(o)}. \tag{9}$$

Based on Eq. (9), $\{\bar{1}\| \bar{1}\}$ operation (i.e. *PT*) causes the signs of all $\sigma_{jk}^{\alpha(o)}$ components to reverse. Consequently, antiferromagnetic materials with *PT* symmetry do not exhibit *T*-odd SHE. Besides, pure spin operations $\{m\|E\}$ in coplanar and collinear antiferromagnets do not switch the signs of all *T*-odd SHE elements. Therefore, unlike the AHE and SHG effects, the collinear and coplanar magnetic materials can possess the *T*-odd SHE in the absence of the SOC effect [29,33]. For example, in the coplanar antiferromagnet Mn₃Sn with spin space group $P^{3^1_{001}}6_3 /^1 m^{\frac{m_\pi}{3}} m^{\frac{m_{2\pi}}{3}} c^m 1$, the three *T*-odd SHE subtensors are

$$\sigma^\alpha = \begin{pmatrix} 0 & \sigma_{xy}^\alpha & 0 \\ \sigma_{xy}^\alpha & 0 & 0 \\ 0 & 0 & 0 \end{pmatrix}, \sigma^\beta = \begin{pmatrix} -\sigma_{xy}^\alpha & 0 & 0 \\ 0 & \sigma_{xy}^\alpha & 0 \\ 0 & 0 & 0 \end{pmatrix}, \sigma^\gamma = \begin{pmatrix} 0 & 0 & 0 \\ 0 & 0 & 0 \\ 0 & 0 & 0 \end{pmatrix}.$$

where $\alpha \to S_x, \beta \to S_y, \gamma \to S_z$. The above result is consistent with that in Ref. [33]. Besides, we can also prove that the *T*-odd SHE of altermagnet MnTe (spin space group $P^{-1}6_3 /^{-1} m^1 m^{-1} c^{\infty m} 1$) vanishes in the absence of the SOC effect, even though its band is spin-splitting [29].

*SPGE tensors under the spin groups.* Using the spin groups, we can also obtain the symmetry-constrained SPGE tensors in the absence of SOC effect. Under the spin group operation $\{U \| R\}$, the *T*-even SPGE tensor $\chi_{ijk}^{\alpha(e)}$ transforms

$$\{U \| R\}: \chi_{ijk}^{\alpha(e)} = \det(U)\sum_{\beta lmn} U_{\alpha\beta} R_{il} R_{jm} R_{kn} \chi_{lmn}^{\beta(e)}, \tag{10}$$

and the transformation of the *T*-odd tensor $\chi_{ijk}^{\alpha(o)}$ is given by

$$\{U \| R\}: \chi_{ijk}^{\alpha(o)} = \sum_{\beta lmn} U_{\alpha\beta} R_{il} R_{jm} R_{kn} \chi_{lmn}^{\beta(o)}. \tag{11}$$

From the above equations, one can see that the *PT* symmetry $\{\bar{1} \parallel \bar{1}\}$ forbids *T*-even but allows *T*-odd SPGE when the SOC effect is absent. For example, the spin group of monolayer of MnPS$_3$ is $P^{-1}\bar{3}^1 1^{-1} m^{\infty m} 1$, and the *T*-even SPGE tensor is zero according to Eq. (10), and the *T*-odd SPGE tensor based on Eq. (11) is:

$$\sigma^{S_z} = \begin{pmatrix} 0 & 0 & 0 & \sigma^{\gamma}_{Xyz} & 0 & \sigma^{\gamma}_{Xxy} \\ \sigma^{\gamma}_{Xxy} & -\sigma^{\gamma}_{Xxy} & 0 & 0 & -\sigma^{\gamma}_{Xyz} & 0 \\ 0 & 0 & 0 & 0 & 0 & 0 \end{pmatrix}.$$

and $\sigma^{S_x} = \sigma^{S_y} = 0$, which is consistent with Ref. [3]. This characteristic leads to the photocurrent being parallel with the magnetic moment all the time in the absence of the SOC effect.

In a short summary, the spin group method can determine the SHE and SPGE tensors, aligning with approaches that utilize spin-resolved electron bands in reciprocal space [29,30] and spin sublattices in real space [3,32]. Notably, the spin group method is versatile, applicable not only to collinear magnetic materials but also to non-collinear magnetic materials.

Since the spin group of a specific magnetic material usually has more symmetry operations than its magnetic group, the symmetry-adapted tensors constrained by the spin group generally have more zero tensor elements than those of magnetic groups (see example in Sect. VI). Projecting the response tensors from the spin space to real space, and comparing the symmetry-constrained tensors under magnetic groups and spin groups, we can disentangle the SOC effect in the response effect, as shown Fig. 2.

**2.3 Response tensors with the magnetic order**

Recently, the relationships between the response tensor and the magnetic order (such as the Néel vector) in magnetically ordered materials [21,38,54-63] have attracted widespread attention. However, the discrete magnetic group and spin group treat the magnetic order as an intrinsic structure parameter, making it difficult to establish a direct connection between response tensors and magnetic orders continuously.

Here, we develop a symmetry method named the "extrinsic parameter" method, where the magnetic order is treated as an extrinsic parameter, to build the analytical relationships between the magnetic orders and response tensors. In the following, we will implement the method to AHE, SHG and SHE for the collinear antiferromagnets.

*AHE vector with magnetic orders*. AHE conductivity tensor $\left[\sigma^A_{ij}\right]$ in Eq. (5) can act as a pseudo-vector $\boldsymbol{\sigma}_H = (\sigma_x, \sigma_y, \sigma_z)$, where $\sigma_i = \varepsilon_{ijk}\sigma^A_{jk}$ ($i,j,k \in (x,y,z)$). We expand $\boldsymbol{\sigma}_H$ as a Taylor polynomial for magnetic order, such as Néel vector **n**:

$$\boldsymbol{\sigma}_H = \mathbf{T}^{(2)} \cdot \mathbf{n} + \mathbf{T}^{(4)} \vdots \mathbf{nnn} + \cdots \quad (12)$$

where the first and third Taylor expression tensor $\mathbf{T}^{(2)}$ and $\mathbf{T}^{(4)}$ are the two- and four-order matrixes, respectively. The even terms on the Néel vector are missing because $\boldsymbol{\sigma}_H$ is odd under the time-reversal operation. Since the magnetic structure has been considered as the extrinsic parameter, only the space group should be utilized to constrain $\left[\sigma_{ij}^A\right]$ and Néel vector $\mathbf{n}$.

The space group operation $\{R|\mathbf{t}\}$ acting on the Néel vector $\mathbf{n}$ results in the change of $\boldsymbol{\sigma}_H$, and it is equivalent to applying the symmetry operation on the AHE conductivity vector $\boldsymbol{\sigma}_H$, i.e.,

$$\boldsymbol{\sigma}_H(\{R|\mathbf{t}\}\mathbf{n}) = R\boldsymbol{\sigma}_H. \tag{13}$$

Here, the translation operation $\mathbf{t}$ is omitted on the right side, as it has no impact on macroscopic $\boldsymbol{\sigma}_H$.

The transformation of $\boldsymbol{\sigma}_H$ under the space group symmetry operation $\{R|\mathbf{t}\}$ in Eq. (13) becomes:

$$\{R|\mathbf{t}\}: \boldsymbol{\sigma}_H \to \det(R)D(R)\boldsymbol{\sigma}_H. \tag{14}$$

The transformation of the Néel vector $\mathbf{n}$ under $\{R|\mathbf{t}\}$ is given by:

$$\{R|\mathbf{t}\}: \mathbf{n} \to \pm\det(R)D(R)\mathbf{n}. \tag{15}$$

Here, the $\pm$ signs indicate whether the two magnetic sublattices with opposite spins are exchanged.

Correspondingly, $\mathbf{T}^{(2)}$ must satisfy the following transformation relationship:

$$T_{ij}^{(2)} = \pm \sum_{mn} R_{im} R_{jn} T_{mn}^{(2)}. \tag{16}$$

Similarly, the transformation of the fourth-order tensor $\mathbf{T}^{(4)}$ under symmetry operation $\{R|\mathbf{t}\}$ is given by:

$$T_{ijkl}^{(4)} = \pm \sum_{mnpq} R_{im} R_{jn} R_{kp} R_{lq} T_{mnpq}^{(4)}. \tag{17}$$

Additionally, $\mathbf{T}^{(4)}$ exhibits commutative symmetry for the last three indices:

$$T_{ijkl}^{(4)} = T_{ijlk}^{(4)} = T_{ikjl}^{(4)} = T_{iklj}^{(4)} = T_{ilkj}^{(4)} = T_{iljk}^{(4)}. \tag{18}$$

By substituting $\mathbf{T}^{(2)}$ and $\mathbf{T}^{(4)}$ into Eq. (12), we can derive the relationship between the AHE vector $\boldsymbol{\sigma}_H$ and Néel vector $\mathbf{n}$ up to the third order. Using the "extrinsic parameter" symmetry method, we have revealed that the AHE vectors $\boldsymbol{\sigma}_H$ in altermagnets can exhibit diverse non-trivial textures in the Néel order space [40].

*T-odd SHE tensors with Néel vectors*. Next, we consider the relationship between the *T*-odd SHE tensor and Néel vector $\mathbf{n}$ in collinear antiferromagnets. Similar to Eq. (12), we can expand a SHE tensor in terms of the Néel vector $\mathbf{n}$. Since the SHE tensors are third-order tensors, their first-order Taylor expansion of Néel vector $\mathbf{n}$ results in a fourth-order polar-like tensor $\mathbf{T}^{(4)}$. It is transformed under the space group operation $\{R|\mathbf{t}\}$ as:

$$\{R|\mathbf{t}\}: T_{ijK}^{\alpha} = \pm \sum_{\beta abC} R_{\alpha\beta} R_{ia} R_{jb} R_{KC} T_{abC}^{\beta}, \tag{19}$$

where the superscripts here denote spin directions, the lowercase English letters in subscript indicate the directions of electric fields or currents, and the capital letters represent the directions of the Néel vector.

Furthermore, the coefficients of the third-order Taylor expansion constitute a sixth-order tensor $\mathbf{T}^{(6)}$, which is transformed under the space group operation $\{R|\mathbf{t}\}$ as:

$$\{R|\mathbf{t}\}: T^{\alpha}_{ijKLM} = \pm \sum_{\beta abCDE} R_{\alpha\beta} R_{ia} R_{jb} R_{KC} R_{LD} R_{ME} T^{\beta}_{abCDE}. \tag{20}$$

Additionally, the capital subscript indices "CDE" in $T^{\beta}_{abCDE}$, representing the direction of the Néel vector $\mathbf{n}$, possess exchange symmetry. Finally, considering both the first-order and third-order Taylor expansions of the Néel vector, the $T$-odd SHE tensor can be expressed as:

$$\sigma^{\alpha}_{ab}(\mathbf{n}) = \sum_{C=1}^{3} T^{\alpha}_{abC} n_C + \sum_{C,D,E=1}^{3} T^{\alpha}_{abCDE} n_C n_D n_E. \tag{21}$$

*T-odd SHG tensors with Néel vectors.* Next, let's consider the relationship between $T$-odd SHG tensor and the Néel vector. Since a set of SHG coefficients form a third-order tensor, its first-order Taylor expansions on the Néel vector result in a fourth-order tensor, satisfying the following relationship:

$$\{R|\mathbf{t}\}: T^{(4)}_{ijkL} = \pm \sum_{lmnP} \det(R) R_{il} R_{jm} R_{kn} R_{LP} T_{lmnP}. \tag{22}$$

Unlike Eq. (19), the term $\det(R)$ cannot be canceled out. Similarly, the third-order Taylor expansion of the $T$-odd SHG tensor constitutes a sixth-order tensor $\mathbf{T}^{(6)}$, and the symmetry constraints on this tensor are:

$$\{R|\mathbf{t}\}: T^{(6)}_{ijkLMN} = \pm \sum_{abcDEF} \det(R) R_{ia} R_{jb} R_{kc} R_{LD} R_{ME} R_{NF} T^{(6)}_{abcDEF}. \tag{23}$$

Additionally, we need to consider the exchange symmetry of the two indices $\{j, k\}$ / $\{b, c\}$ for the electric field and three indices of $\{L, M, N\}$ / $\{D, E, F\}$ for the Néel vector. Finally, considering both the first-order and third-order Taylor expansions of Néel vector $\mathbf{n}$, the SHG coefficients can ultimately be expressed as:

$$\chi_{ijk}(\mathbf{n}) = \sum_{L=1}^{3} T^{(4)}_{ijkL} n_L + \sum_{L,M,N=1}^{3} T^{(6)}_{ijkLMN} n_L n_M n_N. \tag{24}$$

It is worth emphasizing here that, despite the absence of an explicit time-reversal symmetry operation in the extrinsic parameter method, the ±1 symbols in the above equations essentially reflect the change of the magnetic moment under the time-reversal operation. Consequently, the response tensors obtained through the extrinsic parameter method align with the results from the magnetic group and spin group methods (see examples in Sec. VI). Furthermore, the extrinsic parameter method facilitates a clear connection between the response tensors and magnetic orders. More importantly, this method establishes the analytical expression on response tensors in the magnetic order space, providing a decoding table that translates the response tensors into the magnetic orders for Néel vectors detection.

## 3. Code structure and usage of TensorSymmetry

Despite the significant theoretical differences among the above three methods, the approaches to derive the invariant tensor elements are basically the same. Given Wolfram Mathematica's expertise in symbolic computations and its human-friendly output format, we developed a computational program on this software named "TensorSymmetry". The workflow of this program is depicted in Fig. 3. The package can investigate the response tensors constrained by magnetic point groups and spin point groups to disentangle the SOC effect, and establish the analytical relationship of response tensors and magnetic moments with the "extrinsic parameter" method.

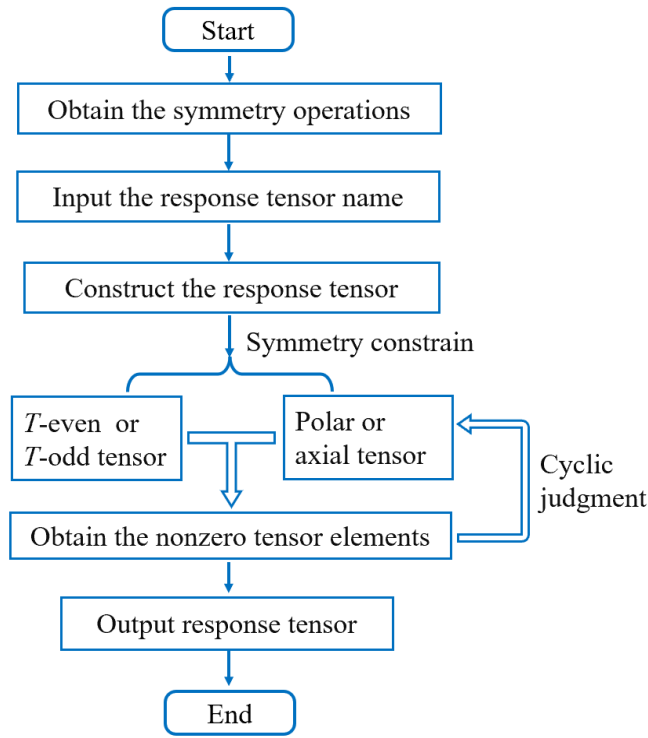

Fig. 3 The workflow of TensorSymmetry.

The software package consists of four kinds of modules:
(1) *SGData.wl, MPGData.wl*: These modules contain all the symmetry operations for space groups and magnetic point groups, respectively.
(2) *TensorSymmetryMG.wl*: This module is to calculate the response tensors constrained by the magnetic groups.
(3) *TensorSymmetrySG.wl*: This module is to study the response tensors under the spin groups.
(4) *TensorSymmetryEP.wl*: This module is to establish the analytical relationships between the response tensors and the magnetic orders.

Here, we introduce the usage of the TensorSymmetry briefly.
   A. *Response tensors constrained by magnetic groups.*
13

(1) Open the user interface file "ToUseMagneticGroup.nb". This file should be put in the same directory as the script folder.

(2) Enter the magnetic point group number, and use the "MPGmassage" function to get the symmetry operation matrix (MPGop) and basis vectors (Basis), as follows:

```
MPGno={14,3};
{MPGop,Basis}=MPGmassage[MPGno];
```

{14,3} is the number of magnetic point group -4'2'm. We can refer to "Point Group Tables" (https://www.cryst.ehu.es/cryst/mpoint_uni.html) to find the numbers of all magnetic point groups. For a magnetic material, one can use FINDSYM (https://stokes.byu.edu/iso/findsym.php) or a related website to determine its magnetic space group and corresponding magnetic point group.

(3) Input the desired tensor function and calculate it.

```
ConductivityTensor[MPGop, Basis];   (*Calculate the normal conductivity and anomalous Hall conductivity tensor*)
SHGtensor[MPGop, Basis];   (*Calculate the SHG tensor*)
SPGEtensor[MPGop, Basis];   (*Calculate the spin photogalvanic effect tensor*)
SHEtensor[MPGop, Basis];   (*Calculate the spin Hall tensor*)
AxialTensor[MPGop, Basis];   (*Calculate the axial tensor, such as CPGE, Edelstein effect, magnetoelectric effect*)
```

The independent tensor components and their number will be printed, as shown in Fig. 4.

```
MPGno = {18, 2};
{MPGop, Basis} = MPGmassage[MPGno];
SHGtensor[MPGop, Basis];

The name of the Magnetic point group: 321'

Independent elements of even SHG tensor: {χ_112, χ_123}, Number: 2
```

$$\begin{pmatrix} 0 & 0 & 0 & \chi_{123} & 0 & \chi_{112} \\ \chi_{112} & -\chi_{112} & 0 & 0 & -\chi_{123} & 0 \\ 0 & 0 & 0 & 0 & 0 & 0 \end{pmatrix}$$

Independent tensor elements of odd SHG tensor: {}, Number: 0

$$\begin{pmatrix} 0 & 0 & 0 & 0 & 0 & 0 \\ 0 & 0 & 0 & 0 & 0 & 0 \\ 0 & 0 & 0 & 0 & 0 & 0 \end{pmatrix}$$

Fig. 4   The input and output of *ToUseMagneticGroup.nb* to determine the symmetry-constrained response tensor under the magnetic point group.

We also provide a dictionary on dielectric tensors, AHE tensors, SHG tensors, SHE tensors, and SPGE tensors for all magnetic point groups. These can be found on the website https://ruichun.github.io/TensorSymmetry/.

B. *Response tensors constrained by the spin group.*

Since there is no comprehensive operation database on spin groups, we use third-party tools like FINDSPINGROUP (https://findspingroup.com/) [64] to determine the spin group operations for specific magnetically ordered materials. Then, the user needs to manually input the generator operation elements $\{U\|R\}$ in *TensorSymmetrySG.wl*. Nevertheless, all the possible matrices of *U* and *R* are already given in our program.

Then, with these matrices, we construct the generator operations for a specific spin group. For example,

```
SGop1={S2,C4z};   (*The first of generator operation of spin group operation in {U||R} form*)
SGop2={C1,mz};   (*The second of generator operation of spin group operation*)
SGop3={S2,mx};
SGop4={C1,mxy};
SGop5={C∞,C1};
SGop6={mx,C1};
RuO2SG={SGop1, SGop2, SGop3, SGop4, SGop5, SGop6};
SpinGroupSHE[RuO2SG]
SpinGroupSHG[RuO2SG]
SpinGroupSPGE[RuO2SG]
SpinGroupSHE[RuO2SG]
```

*C. Response tensor with the magnetic order.*

We should input the space group number and the positions of magnetic atoms with spin-up and spin-down states. For example,

```
{spgop, Basis}=SpaceGroupData[136]; (*136 is the space group number of RuO2*)
Magup={{0,0,0}};   (*Magnetic atom position with spin up state, in crystal coordinate*)
Magdn={{1/2,1/2,1/2}};   (*Magnetic atom position with spin-down state, in crystal coordinate*)
AHEdirection[spgop, Basis, Magup];   (*AHE tensor with the Néel vector *)
SHEdirection[spgop, Basis, Magup];   (*SHE tensor with the Néel vector *)
SHGdirection[spgop, Basis, Magup];   (* SHG tensor with the Néel vector*)
```

Additionally, our program also has the capability to analyze response tensors with magmatic moment order for collinear ferromagnets. The users simply need to configure all magnetic atoms in the "Magup" array while leaving "Magdn" as an empty set.

## 4. Examples

In the following, we present two examples to show how to disentangle the SOC effect and establish the analytical relationships between the response tensors and magnetic orders via the three methods.

*The T-odd SHE of rutile-type altermagnets*. In altermagnets, the *T*-odd SHE can occur in the absence of the SOC effect [29]. Altermagnets with a rutile structure, such as $RuO_2$ and $MnF_2$, belong to the spin space group $P^{-1}4_2/^1m^{-1}n^1m^{\infty m}1$. With the spin group, we can calculate the symmetry-constrained *T*-odd SHE tensor, and it is presented in Table III, which is consistent with Ref. [29]. Notably, only the *T*-odd SHE conductivity $\sigma^z$ exists, resulting in the spin polarization of the spin current $\mathbf{J}^\gamma$ always being parallel to the spin direction of the magnetic moment. This is quite different from the *T*-even SHE in non-magnetic materials, where the spin polarization direction is often perpendicular to the charge current due to the SOC effect.

When the SOC effect is considered, the magnetic point groups should be utilized to seek the symmetry-constrained SHE tensors. If the magnetic moments of these rutile altermagnets are aligned along the *z*-direction, the magnetic point group is 4'/mmm', the *T*-odd SHE tensors are listed in Table III. The additional nonzero $\sigma^X$ and $\sigma^Y$ subtensors arise from the SOC effect. Similarly, when the magnetic moments is along the *x*- (*y*-) direction, the magnetic group belongs to m'mm' (mm'm'), the main subtensor elements are $\sigma^X$ ($\sigma^Y$), while the nonzero $\sigma^Y$ and $\sigma^Z$ ($\sigma^X$ and $\sigma^Z$) tensor elements arise from the SOC effect.

The SHE tensor with the Néel vector using the extrinsic parameter method up to the first order is also listed in Table III, which aligns with the results obtained from the magnetic point groups, and more information along the general magnetic direction can also be obtained.

Table III. *T*-odd SHE tensor for rutile-type antiferromagnets using three methods. The SHE coefficients are expanded to the first-order Taylor expansion of the Néel vector.

| Method | Group | $\sigma^X$ | $\sigma^Y$ | $\sigma^Z$ |
|---|---|---|---|---|
| Spin point group | ${}^{-1}4/{}^1m^{-1}m^1m^{\infty m}1$ | $\begin{pmatrix} 0 & 0 & 0 \\ 0 & 0 & 0 \\ 0 & 0 & 0 \end{pmatrix}$ | $\begin{pmatrix} 0 & 0 & 0 \\ 0 & 0 & 0 \\ 0 & 0 & 0 \end{pmatrix}$ | $\begin{pmatrix} 0 & \sigma^\gamma_{xy} & 0 \\ \sigma^\gamma_{xy} & 0 & 0 \\ 0 & 0 & 0 \end{pmatrix}$ |
| | Projecting to real space | $\sigma^\gamma_{xy}\begin{pmatrix} 0 & n_x & 0 \\ n_x & 0 & 0 \\ 0 & 0 & 0 \end{pmatrix}$ | $\sigma^\gamma_{xy}\begin{pmatrix} 0 & n_y & 0 \\ n_y & 0 & 0 \\ 0 & 0 & 0 \end{pmatrix}$ | $\sigma^\gamma_{xy}\begin{pmatrix} 0 & n_z & 0 \\ n_z & 0 & 0 \\ 0 & 0 & 0 \end{pmatrix}$ |
| Magnetic point group | [001]: 4'/mnm' | $\begin{pmatrix} 0 & 0 & 0 \\ 0 & 0 & \sigma^x_{yz} \\ 0 & \sigma^x_{zy} & 0 \end{pmatrix}$ | $\begin{pmatrix} 0 & 0 & \sigma^x_{yz} \\ 0 & 0 & 0 \\ \sigma^x_{zy} & 0 & 0 \end{pmatrix}$ | $\begin{pmatrix} 0 & \sigma^z_{xy} & 0 \\ \sigma^z_{xy} & 0 & 0 \\ 0 & 0 & 0 \end{pmatrix}$ |
| | [100]: m'mm' | $\begin{pmatrix} 0 & \sigma^x_{xy} & 0 \\ \sigma^x_{yx} & 0 & 0 \\ 0 & 0 & 0 \end{pmatrix}$ | $\begin{pmatrix} \sigma^y_{xx} & 0 & 0 \\ 0 & \sigma^y_{yy} & 0 \\ 0 & 0 & \sigma^y_{zz} \end{pmatrix}$ | $\begin{pmatrix} 0 & 0 & 0 \\ 0 & 0 & \sigma^z_{yz} \\ 0 & \sigma^z_{zy} & 0 \end{pmatrix}$ |
| | [010]: mm'm' | $\begin{pmatrix} \sigma^x_{xx} & 0 & 0 \\ 0 & \sigma^x_{yy} & 0 \\ 0 & 0 & \sigma^x_{zz} \end{pmatrix}$ | $\begin{pmatrix} 0 & \sigma^y_{xy} & 0 \\ \sigma^y_{yx} & 0 & 0 \\ 0 & 0 & 0 \end{pmatrix}$ | $\begin{pmatrix} 0 & 0 & \sigma^z_{xz} \\ 0 & 0 & 0 \\ \sigma^z_{zx} & 0 & 0 \end{pmatrix}$ |
| Extrinsic parameter method | P4$_2$/mnm | $\begin{pmatrix} T_{Xxxy}n_y & T_{Xxyx}n_x & 0 \\ T_{Xyxx}n_x & T_{Xyyy}n_y & T_{Xyzz}n_z \\ 0 & T_{Xzyz}n_z & T_{Xzzy}n_y \end{pmatrix}$ | $\begin{pmatrix} T_{Xyyy}n_x & T_{Xyxx}n_y & T_{Xyzz}n_z \\ T_{Xxyy}n_y & T_{Xxxy}n_x & 0 \\ T_{Xzyz}n_z & 0 & T_{Xzzy}n_x \end{pmatrix}$ | $\begin{pmatrix} 0 & T_{Zxyz}n_z & T_{Zzzy}n_y \\ T_{Zxyz}n_z & 0 & T_{Zzzy}n_x \\ T_{Zzxy}n_y & n_xT_{Zzxy} & 0 \end{pmatrix}$ |

*SHG in monolayer MnPSe$_3$.* The monolayer Néel-type antiferromagnetic material MnPSe$_3$ possesses *PT* symmetry, and it is an XY magnet whose Néel vector can be effectively controlled by strain [39]. Accordingly, the *T*-even SHG is zero, while the *T*-odd SHG exists and is determined by the direction of the magnetic moment. Therefore, SHG can be employed to detect the Néel vector [39]. Regarding its spin point group $^{-1}\bar{3}^11^{-1}m^{\infty m}1$, *T*-odd SHG is zero, indicating that the SOC is necessary for the SHG effect. The *T*-odd SHG tensors with different magnetic directions with the magnetic group method are listed in Table IV. The results are consistent with those in Ref. [36,65]. The analytical relationship between the SHG tensor and the Néel vector can be obtained according to the fact that its parent space group of $P\bar{3}1m$, and two spin opposite magnetic atoms are occupied at (1/3, 2/3, 0) and (2/3, 1/3, 0), which is presented in Table IV. The SHG tensor obtained through the extrinsic parameter method aligns with those derived from magnetic groups. Furthermore, the analytical relationship can be established, which proves useful for analyzing the SHG behavior when the Néel vector evolves continuously.

If the magnetic moment lies in the *xy*-plane, the six in-plane SHG coefficients $\chi_{11}$, $\chi_{12}$, $\chi_{16}$, $\chi_{21}$, $\chi_{22}$ and $\chi_{26}$ with first-order Néel order are given by

$$\begin{cases} \chi_{11} = T_{Xxxx}n_x, \\ \chi_{12} = (T_{Xxxx} - 2T_{Xxyy})n_x, \\ \chi_{26} = T_{Xxyy}n_x, \end{cases} \begin{cases} \chi_{16} = T_{Xxyy}n_y, \\ \chi_{21} = (T_{Xxxx} - 2T_{Xxyy})n_y, \\ \chi_{22} = T_{Xxxx}n_y. \end{cases} \quad (25)$$

Utilizing the above relationship, we can obtain the polarization-resolved SHG patterns as follows

$$\begin{cases} I_{//} = T_{Xxxx}^2 \cos(\varsigma - \theta)^2, \\ I_{\perp} = (T_{Xxxx} - 2T_{Xxyy})^2 \sin^2(\varsigma - \theta), \end{cases} \quad (26)$$

where $\varsigma$ ($\theta$) is the angle between the Néel vector (polarization light) and the *x-axis*. The parallel ($I_{\perp}$) and crossed ($I_{//}$) SHG pattern correspond to E(2ω) ∥ E(ω) and E(2ω) ⊥ E(ω), respectively. Eq. (26) indicates the $I_{\perp}$ and $I_{//}$ show the two-fold rotation symmetry. It can clearly be seen that the $I_{\perp}$ has the minimum valve at $\varsigma = \theta$, as shown in Fig. 5, which is consistent with the experiments in Ref. [39].

Table IV. The relationship of *c*-type SHG coefficients of monolayer MnPSe$_3$ with Néel vector.

The SHE coefficients is expanded to the first-order Taylor expansion of the Néel vector.

| Method | Group | *T*-odd SHG tensor |
|---|---|---|
| Spin group | $^{-1}\bar{3}^1 1^{-1} m^{\infty m} 1$ | 0 |
| Magnetic point group | [001]: $\bar{3}'1m'$ | $\begin{pmatrix} \chi_{11} & -\chi_{11} & 0 & 0 & \chi_{15} & 0 \\ 0 & 0 & 0 & \chi_{15} & 0 & -\chi_{11} \\ 0 & 0 & 0 & 0 & 0 & 0 \end{pmatrix}$ |
| Magnetic point group | [100]: 2'/m | $\begin{pmatrix} \chi_{11} & \chi_{12} & \chi_{13} & 0 & \chi_{15} & 0 \\ 0 & 0 & 0 & \chi_{24} & 0 & \chi_{26} \\ \chi_{31} & \chi_{32} & \chi_{33} & 0 & \chi_{35} & 0 \end{pmatrix}$ |
| Magnetic point group | [120]: 2/m' | $\begin{pmatrix} 0 & 0 & 0 & \chi_{14} & 0 & \chi_{16} \\ \chi_{21} & \chi_{22} & \chi_{23} & 0 & \chi_{25} & 0 \\ 0 & 0 & 0 & \chi_{34} & 0 & \chi_{36} \end{pmatrix}$ |
| Extrinsic parameter method | $P\bar{3}1m$ | $\begin{pmatrix} T_{Xxxx}n_x + T_{Xxxz}n_z & (T_{Xxxx} - 2T_{Xxyy})n_x - T_{Xxxz}n_z & T_{Xxxx}n_x & -T_{Xxxx}n_y & T_{Xxxx}n_x + T_{Xxzz}n_z & T_{Xxyy}n_y \\ (T_{Xxxx} - 2T_{Xxyy})n_y & T_{Xxxx}n_y & T_{Xxxx}n_y & T_{Xxzz}n_z - T_{Xxxx}n_x & -T_{Xxxx}n_y & T_{Xxyy}n_x - T_{Xxxx}n_z \\ T_{Zxxx}n_x + T_{Zxxz}n_z & T_{Zxxx}n_z - T_{Zxxx}n_x & T_{Zxzz}n_z & T_{Zxxx}n_y & T_{Zxzx}n_x & -T_{Zxxx}n_y \end{pmatrix}$ |

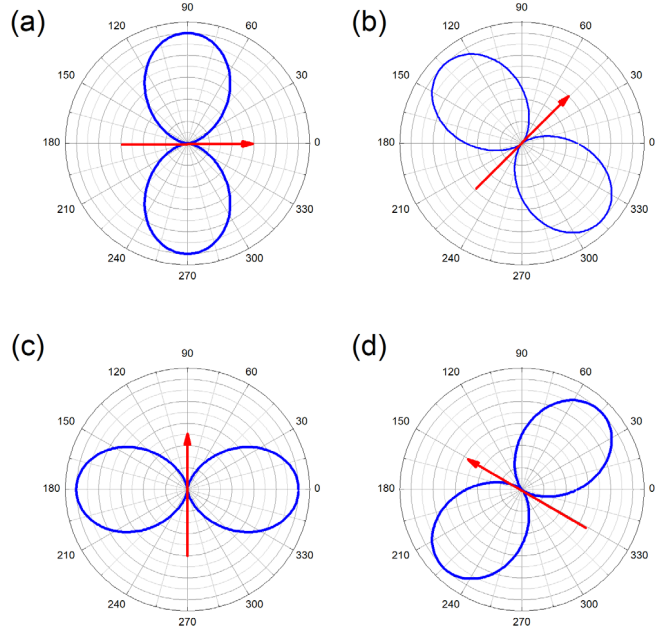

Fig. 5. In-plane polarization-resolved crossed ($I_\perp$) SHG signals (blue) with the Néel vector (red) with $\varsigma$ = (a) 0, (b) 45°, (c) 90° and (d) 135° for MnPSe$_3$.

## 5. Conclusion

In this work, we developed a comprehensive platform "TensorSymmetry", which includes the magnetic group, spin group, and extrinsic parameter methods, to determine the symmetry-adapted tensors in magnetic materials. Our work provides an integrated tool to disentangle the SOC effect and establish the analytical relationship with magnetic order, which offers valuable guidance for experimental detection and theoretical calculations.

# Acknowledgment

We thank the useful discussion with Ze-Ying Zhang and Fang-Chu Chen. We thank the support from the National Natural Science Foundation of China (Grants Nos. 12474100, 12204009), and the Natural Science Foundation of Anhui Province (Grant No. 2208085QA08).